\documentclass[sigconf]{acmart}

\pdfoutput=1
\usepackage{booktabs} 

\usepackage{caption}
\usepackage{subcaption}
\usepackage{makecell}
\usepackage{amsmath}
\usepackage{bm}
\usepackage{amssymb}
\usepackage{multirow}

\newcommand*\samethanks[1][\value{authornote}]{\authornotemark[#1]}

\setcopyright{rightsretained}

\acmDOI{10.475/123_4}

\acmISBN{123-4567-24-567/08/06}

\copyrightyear{2018} 
\acmYear{2018} 
\setcopyright{acmcopyright}
\acmConference{WSDM 2019}{February, 2019}{Melbourne, Australia}\acmPrice{15.00}\acmDOI{10.1145/3125486.3125491}

\begin{document}
\title{Time is of the Essence: a Joint Hierarchical RNN and Point Process Model for Time and Item Predictions}
\titlenote{This work was carried out at the Telenor-NTNU AI-Lab, hosted by the Department of Computer Science, Norwegian University of Science and Technology.}

\author{Bj{\o}rnar Vass{\o}y}
\authornote{All the authors contributed equally}
\affiliation{%
  \institution{Department of Computer Science, Norwegian University of Science and Technology}
}
\email{bjorva@stud.ntnu.no}

\author{Massimiliano Ruocco}
\samethanks[2]
\authornote{This author is also affiliated with Telenor Research.}
\affiliation{%
  \institution{Department of Computer Science, Norwegian University of Science and Technology}
}
\email{massimiliano.ruocco@ntnu.no}

\author{Eliezer de Souza da Silva}
\samethanks[2]
\affiliation{%
  \institution{Department of Computer Science, Norwegian University of Science and Technology}
}
\email{eliezer.souza.silva@ntnu.no}

\author{Erlend Aune}
\samethanks[2]
\affiliation{%
  \institution{Exabel AS}
}
\email{aune@exabel.com}

\renewcommand{\shortauthors}{M. Ruocco et al.}

\begin{abstract}
In recent years session-based recommendation has emerged as an increasingly applicable type of recommendation. As sessions consist of sequences of events, this type of recommendation is a natural fit for Recurrent Neural Networks (RNNs). Several additions have been proposed for extending such models in order to handle specific problems or data. Two such extensions are 1.) modeling of inter-session relations for catching long term dependencies over user sessions, and 2.) modeling temporal aspects of user-item interactions. The former allows the session-based recommendation to utilize extended session history and inter-session information when providing new recommendations. The latter has been used to both provide state-of-the-art predictions for when the user will return to the service and also for improving recommendations. In this work we combine these two extensions in a joint model for the tasks of recommendation and return-time prediction. The model consists of a Hierarchical RNN for the inter-session and intra-session items recommendation extended with a Point Process model for the time-gaps between the sessions. The experimental results indicate that the proposed model improves recommendations significantly on two datasets over a strong baseline, while simultaneously improving return-time predictions over a baseline return-time prediction model.
\end{abstract}

%
%

\copyrightyear{2019}
\acmYear{2019}
\setcopyright{acmlicensed}
\acmConference[WSDM '19]{The Twelfth ACM International Conference on
Web Search and Data Mining}{February 11--15, 2019}{Melbourne, VIC,
Australia}
\acmBooktitle{The Twelfth ACM International Conference on Web Search and
Data Mining (WSDM '19), February 11--15, 2019, Melbourne, VIC, Australia}
\acmPrice{15.00}
\acmDOI{10.1145/3289600.3290987}
\acmISBN{978-1-4503-5940-5/19/02}





\maketitle
\fancyhead{} 
\section{Introduction}
\label{sec:introduction}
Finding relevant information on the web is an increasingly challenging problem, as more and more information becomes available. Users trying to find a particular piece of content are likely to enter a state of \textit{information overload}, where they are not able to interpret the available information efficiently. Today, recommender systems are applied ubiquitously to assist users in navigating large sets of web content (video, audio, books, products, etc). A guiding principle in those systems is to apply predictive modeling to infer users' preferences using their history of interactions with the system. In many cases, however, it is not possible to track a long history of interactions, instead, there is only information about recent \textit{sessions} of interactions. Due to increased focus on data privacy over the last years, this situation is likely to become more prevalent. Session-based recommender systems address this issue by being capable of using information available in the sessions themselves. In particular, the predictive problem is cast as a \textit{sequence prediction problem}, where a history with a sequence of actions and interactions taken by the users in a set of sessions are applied to predict the next $k$ actions in a given session or a new session~\cite{hidasi:_session-based_rnn}.

\begin{figure*}[ht!]
 \centering
        \includegraphics[width=0.79\linewidth]{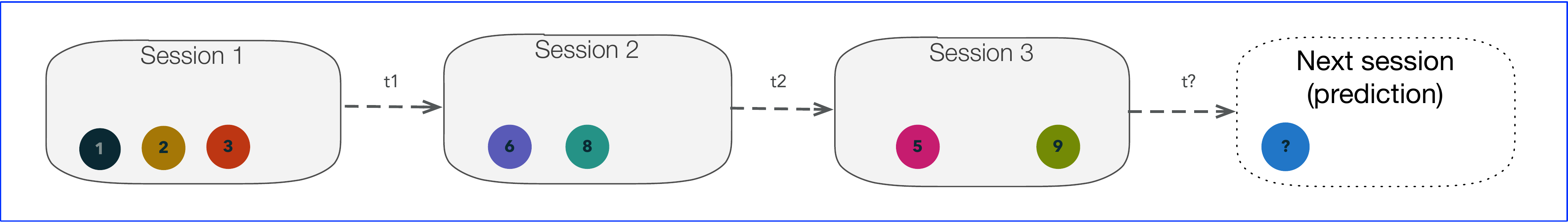}

    \caption{Representation of multiple sequential sessions with sequential item clicks and time between sessions}
    \label{fig:intro_graph}
\end{figure*}

Return time prediction is another important challenge that has been explored using different strategies. The main challenge here is developing a time prediction model that uses previous sessions information to infer \textit{when} the user is going to return or start a new session in a given application or service. This task is important for two reasons: 1.) Modeling user retention rate in web-services through their return time dynamics can offer deep insight into the service at hand, allowing the service decision-makers to better optimize their service to maximize time spent on the service. This is a valuable metric within web economy, which is fueled by advertisement~\cite{DBLP:conf/kdd/KapoorSSY14}. 2.) In the context of recommender systems, there is an interplay between temporal information and recommendations that can be exploited to harness predictive models and better understanding of users engagement with the recommender system~\cite{DBLP:conf/cikm/WuWHS17}.

In this article, we tackle both tasks jointly, based on the assumption that the interplay between users engagement with items and the temporal information is favorable in delivering good predictors for both tasks. We propose a new neural model capable of delivering next--item predictions using Recurrent Neural Networks (RNNs) and return--time predictions using Point Processes. For example, consider the context represented in Figure~\ref{fig:intro_graph}, where there are a sequence of sessions for a given user and within each session a series on user--item interactions (clicks, purchases, etc) with a series of items (represented by the colored circles). Here, there is also information about the time before the user returns to the service and start a new session. Given the motivation given above, it is valuable to build a model with capabilities for capturing inter- and intra sessions user--item dynamics and using the past user--item dynamics and the temporal information to predict when a new session would start, as well as deliver recommendations. 

To summarize, the main contributions of this work are:
\begin{itemize}
\item Defining and introducing a Temporal Hierarchical Recurrent Neural Network
(THRNN): a joint model for inter-session and intra-session recommendations and return-time prediction. It consists in three major components, a module for sequence of sessions of representations, a module for sequences of user-item interactions within each session, a module for the time interval between the sessions. The first two modules consists on a Hierarchical Recurrent Neural Network (HRNN) architecture and the third module is given by a Point Process which shares representations with the HRNN.
\item A tuning mechanism in the training process that allows the time model to modulate the focus on short, medium or long time prediction. The mechanism consists of adding a control parameter in the loss function that allows us to control the importance of temporal information at training time.
\item Experimental evaluation of the proposed model performance in both recommendation and return-time prediction tasks. The results show improvements in both tasks when compared to strong baseline models, and indicate that the joint modeling is particularly beneficial for the return--time prediction task. Finally for the return--time prediction task, we show that the control mechanism for short, medium or long time is effective.
\end{itemize}

\subsection{Problem formulation}\label{subsec:form}

    Given the set of user--item interactions; $S_{\text{train}}=\{(u,i,j,t_j) | u,i,j \in \mathbb{N}, t \in \mathbb{R}^+\}$, each tuple meaning that the user $u$ interacted with the item $i$ in a session $j$ at time $t_j$. We aim to learn a function that estimates a score for new items within a session, a score for an initial recommendation in the next session, and the time gap prediction between the current and next session. More formally, given our training set $S_{\text{train}}$, and an index $j$ for a position within the session, we want a function $f(S_{\text{train}},j)$ that will output a tuple $(s_{j+1},f_{j+1},g_{t_{j+1}})$, with the intra-session recommendation score $s_{j+1}$, the next session initial item score $f_{j+1}$ and the time-gap $g_{t_{j+1}}$.
    
    Thus the objective is to build a model that can predict new items within a session and for the next session, as well as predict the time-gap for the next session (return--time prediction).

\section{Related Work}
\label{sec:relatedwork}
In recent years, numerous deep learning techniques have been successfully employed for recommendations. In particular the use of RNNs has been shown to be promising for session-based recommendation. 

Session-based recommendation have been historically handled by item-to-item/ neighborhood-based methods \cite{sarwar:_item_bases_collab} without considering the sequential nature of session data. A natural choice for modeling this sequential nature are RNN-based models such as those given in \cite{hidasi:_session-based_rnn, RuoccoSL17, quadrana:_person_session_, tan:_improved_rnn_,Jannach:2017:RNN:3109859.3109872}.
The GRU2REC architecture presented in \cite{hidasi:_session-based_rnn} is widely credited to be the first to apply a RNN-based recommendation model and achieving state of the art results. The authors assume that the users with attached sessions are anonymous without any user-history, and assume that the sessions are independent of each other. Each session is modeled as a single layer of GRU units followed by a single feed-forward-layer. The GRU units, having recurrent connections, provide the state and captures the temporal dynamics of the sessions, while the feed-forward layer outputs the scores for each item. The model proposed was tested on two different datasets and achieved a 20-30\% gain in the evaluated measures over \cite{rendle:_bpr_}, which is a nearest-neighbor-based model and was presented as the best performing baseline. A systematic comparison of multiple methods for session-based recommendation and the GRU2REC in \cite{hidasi:_session-based_rnn} is presented in \cite{DBLP:journals/corr/abs-1803-09587}. An extension of the GRU2REC model is presented in \cite{RuoccoSL17}, with a second level of RNNs that tries to capture inter-session dynamics. The proposed model is a hierarchical RNN where one level considers inter-session dynamics and the other considers intra-session dynamics. As opposed to assuming fully anonymous users and totally independent sessions, this extension allows some simple and low-cost user history to be considered in order to provide better recommendations for live sessions. The user history is in the form of abstract representations of previous sessions. The inter-session RNN is fed a finite number of the most recent session-representations in chronological order, and the output is used as the initial hidden state of the intra-session RNN. The motivation behind this is to handle the cold-start problem. 
In contrast to \cite{hidasi:_session-based_rnn}, \cite{RuoccoSL17} achieved the best results when adding an embedding layer for the items. Additionally \cite{RuoccoSL17} did not use the interleaved session mini-batching scheme, instead opting for padded sessions. Nor did they sample the output or use any pairwise losses. The work in \cite{quadrana:_person_session_} proposes an architecture that is very similar to the model in \cite{RuoccoSL17}. They also propose an inter-session RNN layer in order to handle the cold-start problem when one has access to a limited user history in the form of previous sessions. The main difference between this hierarchical model and the one presented in \cite{RuoccoSL17} is that \cite{quadrana:_person_session_} only considered a session representation scheme based on the last hidden state of the intra-session RNN. Additionally, these session representations were created by passing the last hidden state to a final single layer feed-forward layer with a hyperbolic tangent activation function. They also experiment with propagating output of the inter-session network to all time-steps in the intra-session RNN, which complicates the model somewhat, but achieve slightly better results for one of the datasets evaluated. 
The hierarchical model proved to be the best performing overall by significant margin. In \cite{tan:_improved_rnn_}, the approach proposed in \cite{hidasi:_session-based_rnn} is improved by proposing a data-augmentation pre-processing step for improving the robustness of the model as well as proposing to output an item embedding, instead of the individual scores for each item, to make the recommendations faster. The work in \cite{Jannach:2017:RNN:3109859.3109872} presents an approach to adapt a nearest neighbor method for session-based recommendation task by finding sessions that are similar to the live session at the first step using an item cosine similarity measure. For predictions they use a weighted sum of nearest neighbour recommendations and RNN recommendations.

Other notable works on using RNNs have focused on how to best handle context- and feature rich input like \cite{liu:_context-aware_,Smirnova:2017:CSM:3125486.3125488,hidasi:_parallel_rnn_}. In \cite{liu:_context-aware_} a simple intra-session RNN model is extended by training and evaluate using different sets of weights based on the context of the input. This context can for instance be time and date of different granularities, location, or weather at the time of an event. In \cite{hidasi:_parallel_rnn_} the focus is to feed feature-rich input in the RNN input. In \cite{Smirnova:2017:CSM:3125486.3125488}, the authors propose a new class of Contextual Recurrent Neural Networks for Recommendation (CRNNs) taking into account contextual information in two different ways: by combining the context embedding and input embedding, similar to \cite{Twardowski:2016:MCI:2959100.2959162} in one case, and by incorporating it in the model dynamics in the other case. Attention mechanisms have also been used within the RNN setting to provide recommendations \cite{DBLP:conf/cikm/LiRCRLM17}. The idea here is to use an attention mechanism over hidden states to provide a better representation from which to predict subsequent items.

Even though RNNs are inherently able to capture some temporal dependencies, this is largely based on the order of the events in a sequence. Aspects like time gap between events or sessions, the season, the year, the weekday, and the time of the day, are all temporal aspects that may influence the ideal recommendation in many domains, but which cannot easily be captured by the order of events alone. There have been attempts in both making time-aware recommendation by directly feeding such information in the RNN for the simple recommendation task \cite{liu:_context-aware_}, as well as modeling time for the task of predicting the return time of the next session \cite{du:_marked_temporal_point_,jing:_neural_survival_,Zhu:2017:NMU:3172077.3172393}.

The model proposed in \cite{du:_marked_temporal_point_} -- based on a single layer RNN -- attempts to predict the return-time of a user in addition to recommend the next item. The time prediction is modeled as a marked point process with intensity function conditioned on the history and the time to the next event. 
Similarly in combining recommendation and time-modeling, the model proposed in \cite{jing:_neural_survival_} use survival analysis to model the time instead of a marked point process. Another difference is that this model is used for next-basket recommendation task, and not just a single sequence of items. Furthermore, the time modeling is on the inter-basket/session time-gaps, and not the time between each selection. It means using a single level RNN for inter-session modeling directly, in contrast to \cite{RuoccoSL17} and \cite{quadrana:_person_session_}, both implementing some form of inter-session modeling in one of the two levels of their hierarchical RNN models. The final model's recommendation capabilities was compared with two factorization based baselines and one based on neural networks. It was shown to outperform all of these on two different datasets. The time prediction was compared with expressive point processes, one of which was a Hawkes point process, and achieved smaller time errors than all of these. Finally, in \cite{Zhu:2017:NMU:3172077.3172393}, starting from the observation RNN units are good at modeling orders of entities, but does not offer any inherent support of time intervals between the entities, they proposed to incorporate the temporal aspect in the LSTM through explicit time gate.

Also, there has been efforts in using Point Process intensity modeling as an inspiration for hybrid Recurrent Neural Networks capable of time modeling. The approach in~\citet{Du2016RMT} consist in creating a new type of recurrent neural unit with two outputs, one for the time model and another one for the general prediction model. The main contribution here is the proposal of a modeling approach for the intensity function using a combination of the hidden layers for long time dependency and recent time prediction, and combination the usual prediction loss with the negative log-likelihood of the Point process based on the intensity function. This results in a model capable of predicting markers and time of the markers, with an intensity function for the time model that is learned via a neural architecture. In~\citet{DBLP:conf/aaai/XiaoYYZC17} a similar approach is taking, modeling the intensity function of a Point process via neural architecture, however in this case two RNN's are employed: one for time-series data, capturing background intensity rate, and another one for event data, capturing long range relationship between events. Finally,~\citet{DBLP:conf/nips/MeiE17} is proposing a continuous time LSTM using Hawkes process as a starting point and making changes in the internal dynamics of the LSTM to achieve that goal. The resulting model is a LSTM with similar properties as the Hawkes process. 


\section{The Joint Model}
\label{sec:jointmodel}

\begin{figure*}[ht]
\begin{center}
  \includegraphics[width=0.75\textwidth]{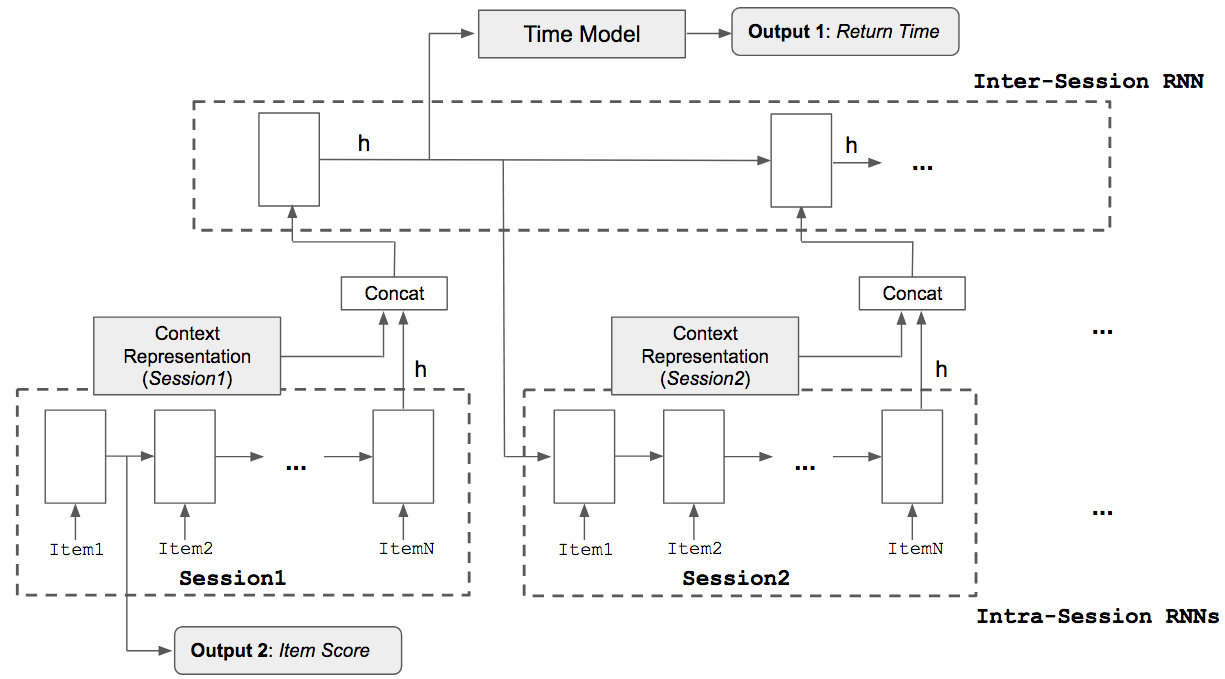}
  \caption{The intra-session RNN}
  \label{fig:proposedmodel}
\end{center}
\end{figure*}

We propose a system that aims to predict the return time for the next session and to recommend the next item. It is based on a hierarchical RNN (HRNN) model enhanced with a time model part. The inputs of the HRNN are the representations of previous sessions together with contextual information, followed by the item representations in the session. Figure \ref{fig:proposedmodel} shows an overview of the system.

The HRNN is inspired by \cite{RuoccoSL17}, and consists of an \textit{inter-session RNN} and an \textit{intra-session RNN}. GRU units are used in both the intra-session and the inter-session RNNs. GRU was chosen due to its ability to remedy the vanishing gradient problem and since it was found to work better than LSTM for this problem and model structure. Like in \cite{RuoccoSL17}, the inter-session RNN is fed with a fixed number of preceding session-representations. Moreover, in the proposed model, relevant session-related contextual information is concatenated to this input.

The final hidden state of the inter-session RNN is propagated to the intra-session RNN mainly representing the inter-session information, as well as employed for the return-time prediction task. The intra-session RNN uses the last hidden state of the inter-session RNN as initial state, and item representations in input. For each item embedding in the input, the corresponding output of the RNN is passed to a linear layer which outputs scores for each target items. The recommendation is given by selecting the items with highest scores.

\subsection{Context Representation}
Additional session-related contextual information, is concatenated with the last hidden state of the intra-session RNN, representing the input of the inter-session RNN. We consider three types of embeddings used in the main setup of the model: 1) Item embeddings, 2) inter-session gap-time embeddings and 3) user embeddings. The purpose of such embeddings is to learn finer dynamics and representations of the embedded entities. For instance, if two different artist often are listened to by users with similar tastes, an embedding layer would most likely learn artist representations that in some sense are similar to each other. By using RNNs, such representations can learn temporal dynamics in addition to a more general "similarity measure". A simplified example of such representational knowledge could be: "Artist A is almost always listened to before Artist B". Embeddings can also learn dissimilarity and non-linearity. For instance when thinking about gap times between sessions, there might be a higher correlation between gap sizes of 24 and 48 hours (periodic daily behavior) than between 12 and 24 hours. 

\subsubsection{Item embeddings}. These embeddings represent each unique item in the dataset.
The learned embeddings are directly handled by the intra-session RNN and consequently only trained by the resulting loss of this part of the model. This means that the item embeddings will affect earlier parts of the network, but then as input, not computational graphs. Hence, their gradient are unable to flow back to the embedding layer, and cannot be used to train the embeddings further. Due to the large number of items present in the considered datasets, the dimensionality of this embedding is the largest by far.

\subsubsection{Inter-session gap-time embedding} These embeddings represent the time gaps between sessions. The time-gap is first normalized and then divided into discrete buckets. The resulting bucket IDs can then be used to index embedding tables/layers to propagate the corresponding embedding. Two different normalization schemes were examined. Both are first given an upper bound, which sets a threshold of the gap time after which we don't consider the user active enough to be provided accurate time predictions. The gap-times that are greater than this bound is set to the upper-bound. The first normalization scheme, divides the gap-time range in uniformly large buckets. The benefit of this is that all the values in the gap-time range will belong to a bucket of equal size, causing no gap-time to be in the same bucket as a much higher or lower gap-time. A disadvantage of using this is that the earlier "popular" buckets can be overcrowded and the later ones can end up being almost empty, which can make such embeddings hard to train. One also needs a high resolution to cover the finer differences in the smaller time-gap ranges, which further increases the problem of sparse buckets. In the second normalization scheme, the gap times are first transformed with a log function, before the transformed range is divided uniformly into buckets. This results in a more evenly distributed number of gap times in the different buckets, but at the cost of cruder resolution for larger gap-times where the corresponding buckets cover much more time than the earlier ones. We observed that the second scheme is performing better with small resolutions, but was overall out-performed by the uniform scheme with higher resolution. Since having high resolution is a non-issue, both with regards to model performance and run-times, the uniform was deemed the better option.

\subsubsection{User embedding} \label{subsec:user_embed}
These embeddings are mainly inspired by \cite{Jing2017NSR}, and learning the user behaviour beyond the history of session representations is especially useful with long user histories. For instance, if a user behaves a bit unusual and sporadic in the last few sessions, a model with user embeddings can have information about long term user behavior, which can help making sense of/override the recent noisy behavior.

\subsection{Time Model}

The main goal of this model is to predict the time until the next session start, given a fixed length history in the form of abstract session representations and corresponding contextual temporal information. The time modeling is heavily inspired by the work in \cite{Du2016RMT}, where time modeling is used to both predict the the time of the next item recommendation given a single sequence of previous items, as well as to improve the recommendation. In their model, the gap-times between selected items are considered to be drawn from a marked point process. The parameterization of the marked point process is defined by the authors and is dependent on the previous selection history modeled as a RNN, with corresponding inter-selection gap-times, and on the time of the last selection. In our system the history is based on session representations and not individual items. Consequently, the corresponding time-gaps are the times between the session representations in the history. This inter-session modeling, as well as the concatenated embeddings, is more similar to what was done in \cite{Jing2017NSR}. In the point process, we define the intensity function as: 
\begin{equation} \label{eq:intesity}
    \lambda^*(t) = \textrm{exp}(v^{t\top}\cdot h_j+w^t\cdot g_j + b^t)
\end{equation}
where $h_j$ is the j-th hidden state, $g_j=t-t_j$ - the time since the last session (in which $t_j$ is the last timestamp in the last session and $t$ is the time variable), $v$ is a vector of weights with the same dimensionality as $h_j$, $w^t$ is a single weight and $b^t$ is a bias term. $v^{t\top}\cdot h_j$ comprises the historical influence on the intensity function, while $w^t\cdot g_j$ is the current influence. 
The full conditional density function is then defined as follows:
\begin{equation} \label{eq:conddens}
    f^*(t) = \lambda^*(t)\textrm{exp}\bigg(-\int_{t_j}^{t} \lambda^*(\tau) d\tau\bigg)
\end{equation}

\noindent finally, where the intensity Equation \ref{eq:intesity} has been substituted into the conditional density function Equation \ref{eq:conddens}, the full expression of the marked point process is:
\begin{multline} \label{eq:finaleq}
    f^*(t) = \textrm{exp}\bigg\{v^{t\top}\cdot h_j+w^t\cdot g_j+b^t+\frac{1}{w^t}\textrm{exp}(v^{t\top}\cdot h_j + b^t)  \\  -\frac{1}{w^t}\textrm{exp}(v^{t\top}\cdot h_j+w^t\cdot g_j + b^t)\bigg\}
\end{multline}



The expected return time $\hat{t}_{j+1}$ is computed as the expected value in form of weighted area over the probability distribution as follows:
\begin{equation} \label{eq:pred}
    \hat{t}_{j+1} = \int_{0}^{\infty}t\cdot f^*(t)dt
\end{equation}
The integration of the density distribution of the point process (Equation \ref{eq:finaleq}) does not have an analytic solution. Thus, prediction has to be approximated by numerical integration. To handle the infinite upper integration bound, a simple upper cut-off time is defined. While this is another approximation, the approximation becomes negligible when setting the cut-off time a bit higher than the vast majority of gap-times found in the data. 

\subsection{Loss}

Creating a loss function is a matter of combining a recommendation loss and a return time loss. The latter is simply the log-likelihood of the time-gap point process distribution. On top of that, in order to control and tune the \textit{importance} of the short, medium and long term data points in the training process, we changed the loss by adding the exponent $\alpha \in (0,1)$ to the time-step variable $g_{j}$. The addition of the exponent on the time-step variable comes from looking at the gradient of the negative log-likelihood in relation to $w^t$: $-\frac{\partial \log f^*(t)}{\partial w^t} = -g_j+\frac{c_1}{(w^t)^2}\textrm{exp}(g_jw^t)(g_jw^t-1)+c_2$ ($c_1$ and $c_2$ are fixed terms not depending directly on the time-interval $g_j$ or the weight $w^t$). The gradient has two basic components related to the time-gap information, a linear component coming from the current session and a exponential component that is related to the past interactions in the Point process model (see discussion of Equation~\ref{eq:intesity}). This led to the conclusion that a simple mechanism to modulate the emphasis between shorter time prediction or longer time prediction is to exponentiate the time-interval variable $g_j$, since this operation would affect the linear and exponential part differently. Another way to see the same intuition is to observe that the negative log-likelihood is approximately linear on small time-gaps and exponential on large time-gaps, meaning that exponentiating the time-gaps to a number closer to zero would make the shorter time-gaps dominate the loss more than the larger time-gaps. This addition proved to be valuable for tuning trade-offs between short-medium-long time predictions as we will see in Section \ref{subsec:results}. The final time loss is:
\begin{multline} \label{eq:timeloss}
    L_{time}(g_{j}, h_j, w) = -\bigg(v^{t\top}\cdot h_j+w^t\cdot g_{j}^{\alpha}+b^t + \\ \frac{1}{w^t}\textrm{exp}(v^{t\top}\cdot h_j + b^t) \\ -\frac{1}{w^t}\textrm{exp}(v^{t\top}\cdot h_j+w^t\cdot g_{j}^{\alpha} + b^t)\bigg)
\end{multline}

The recommendation loss, $L_{rec}(s_{j+1},i)$, is simply the softmax for item $i$ given session representation at time $j+1$, i.e. $s_{j+1}$. We combine these losses using a weighted mean:
\begin{equation} \label{eq:totloss}
    L_{total} = \alpha L_{time}(g_{j}, h_j, w) + \beta L_{rec}(s_{j+1},i)
\end{equation}
Here, $i$ is the target item of the intra-session recommendation. $s_{j+1}$ is the score of the intra-session recommendations and $g_{j}$ is the target time-gap.

\section{Experimental Setting}
\label{sec:experimentalsettings}
In order to ensure the reproducibility of the experiments, and for further implementation details, we make our code available on a github repository \footnote{\url{https://github.com/BjornarVass/Recsys}}.
\subsection{Datasets}
The evaluation is performed on two different datasets as in \cite{RuoccoSL17}: 
the \textit{LastFM} dataset \cite{Bertin-Mahieux2011} containing listening habits of users on the music website Last.fm  and the \textit{Reddit} dataset \footnote{Subreddit interactions dataset: https://www.kaggle.com/colemaclean/subreddit-interactions}, on user activity on the social news aggregation and discussion website Reddit. 
Last.fm is a music website where users can keep track of the songs they listen to as well as sharing this with their peers. The data is in the form of tuples containing \textit{user-id}, \textit{artist}, \textit{song} and \textit{timestamp}, each representing a single listening event. 
Reddit is a popular forum/discussion website, where people can share and comment on different news, creations, pictures and other topics. Its structure is divided into different sub-forums, named subreddits, which define the topics/interests/allegiance of the posts to be posted there. This data is in the form of tuples containing a \textit{user}, a \textit{subreddit} and a \textit{timestamp}, and each of these represents a single event where the user has commented on a post within the specific subforum at the given timestamp.

\subsection{Data Preprocessing}
The data were first preprocessed by removing noisy and irrelevant data and defining markers in the data following the steps in \cite{RuoccoSL17}.
The Reddit dataset contains a log of user interaction on different subreddits (sub-forums), with timestamps. Here, an interaction is when a user adds a comment to a thread. Since the dataset does not split the events into sessions, we did this manually by specifying an inactivity time limit. Using the timestamps, we let consecutive actions that happened within the time limit belong to the same session. That is, for a specified time limit $\Delta_t$, and a list of a user's interactions $\{a_{t_{0}} , a_{t_{1}} , \ldots, a_{t_{n}}\}$, ordered by their timestamps $t_i$, two consecutive interactions $a_{t_{i}}$ and $a_{t_{i+1}}$ belong to the same session if $t_{i+1} \leq t_{i} + \Delta_t$. We set the time limit to $1$ hour ($3600$ seconds) for the LastFM and $30$ minutes ($1800$ seconds) for the Reddit dataset as in \cite{RuoccoSL17}. For both dataset we then removed all $L$ consecutively repeating items, reducing them to only one occurrence. Following \cite{RuoccoSL17} we set the maximum length, $L$, of a session to $L=20$, split sessions that had a length $l>L$ into two sessions and removed sequences of length more than $2L$. As \cite{RuoccoSL17}, we also simplified the LastFM dataset by ignoring the specific song of each user interaction and only use the artists. When modeling the inter-session time-gaps, the sessions that are split because their length are not considered separate sessions, since this would introduce noise. We solved this problem by setting the last timestamp in the first half and the first timestamp in the second half, to be the start time of the full session, resulting in a gap-time equal to $0$. The contribution will then be masked away from the time loss, making sure the model does not to train on these gap-times.
Finally, the datasets were split into a training set and a test set on a per user basis. Each user's sessions were sorted by the timestamp of the earliest event in the session, and the earliest $80\%$ of his sessions were placed in the training set, while the remaining $20\%$ of the sequences, that contain the most recent sessions of each user, were allocated to the test set.
Table \ref{tab:preproc} shows statistics for the two datasets after preprocessing (before splitting into training and test sets).

\begin{table}
  \begin{tabular}{lcc}
    \toprule
     &Reddit&Last.fm\\
    \midrule
    Number of users & 18.271& 977\\
    Number of sessions & 1.135.488& 630.774\\
    Sessions per user & 62,1& 645,6\\
    Average session length & 3,0& 8,1\\
    Number of items & 27.452& 94.284\\
  \bottomrule
\end{tabular}
\caption{Statistics for the datasets after preprocessing}
\label{tab:preproc}
\end{table}

\subsection{Baselines}
In order to validate the performance of our model for both the tasks of return session time prediction and next item recommendation the THRNN model is compared to the following baselines, in addition to the intra-session Hierarchical RNN itself (HRNN). These baselines have been shown to be the strongest ones for such tasks as demonstrated in \citep{RuoccoSL17,quadrana:_person_session_} for the session based recommendation and \cite{du:_marked_temporal_point_} for time prediction.


\paragraph{GRU4REC}\footnote{\url{https://github.com/hidasib/GRU4Rec}}: implementation based on the seminal work of~\citet{hidasi:_session-based_rnn} that uses session-parallel mini-batch training
process and ranking-based loss functions for learning a model using solely the items sequence information within a session.

\paragraph{HRNN}
The Hierarchical RNN (also Inter- and Intra-Session RNN), is a simpler version of the architecture used in this article. Namely, take the architecture presented in Figure \ref{fig:proposedmodel}, remove the time  model and the user embeddings. The resulting model is the one given in \cite{RuoccoSL17}. The HRNN baseline parametrization is the same as in this article where it has been shown to otuperform the GRU4REC and other strongest baselines, such as Item K-NN \cite{Linden:2003:ARI:642462.642471} and BPR-MF \cite{Rendle:2009:BBP:1795114.1795167}

\paragraph{Hawkes process} The Hawkes process is a self-exciting Point process with intensity given by 
\begin{equation}
\lambda(t|\mathcal{H}_t) = \gamma_0 + \alpha \sum_{t_j \in \mathcal{H}_t}\gamma(t,t_j)
\end{equation}
for some constants $\gamma_0, \alpha$, a kernel $\gamma$  and the history of events $\mathcal{H}_t$~\cite{hawkes1971spectra}. $\gamma_0$ defines a baseline intensity, while $\gamma$ defines the propensity with which the intensity is excited by events in the process itself. The kernel typically has the property that if a number of events happen with short time gaps, the intensity $\lambda$ becomes larger than if the same number of events happen with larger gaps. The history of event $\mathcal{H}_t$ has the list of event times up to time $t$, $\{t_1, ... , t_n | \forall j \leq n:t_j < t  \}$. For the baseline, we fit one Hawkes process for each user (using the last 15 sessions), and set the kernel to an exponential function. Return-time predictions are expected values of the Hawkes process primed by the time gaps of the 15 most recent sessions for each user. We also used a \textit{Hawkes long-term} fitting procedure, for which the entire per-user training dataset is used for estimating the parameters in the Hawkes process.


\subsection{Evaluation Metrics and Hyper-Parameters tuning}
We used \textit{Recall@$k$} and \textit{MRR@$k$} with $k = 5,10,20$ to evaluate all models for the recommendation task and \textit{Mean Absolute Error} (MAE), for the return-time prediction. In addition to the baselines already discussed, we also compared the THRNN with other models on the two presented datasets. We experimented with mini-batch sizes, embedding sizes, learning rate, dropout rate, using multiple GRU layers, and number of session representations to find the best configurations for each dataset. 
The best configurations we found are summarized in Table \ref{tab:hyperparam}. \textit{Learning-rate time} refers to the learning rate of $w$ and $v^{t}$ respectively, in the Equation \ref{eq:timeloss}. This value had to be reduced in order to stop this loss from diverging, mainly due to the exponential function as well as the scaling with the time-gap in the formula.  We then find the best scaling factors for the loss function in Equation \ref{eq:totloss} to be $\alpha = 0.45$ and $\beta = 0.45$. 

\begin{table}
  \begin{tabular}{lcc}
    \toprule
     &Reddit&Last.fm\\
    \midrule
    Item Embedding size & 50& 100\\
    User Embedding Size & 10& 10\\
    Time-Gap Embedding Size & 5& 5\\
    Learning rate & 0.001& 0.001\\
    Learning rate-time & 0.0001& 0.0001\\
    Dropout rate & 0& 0.2\\
    Max. recent session representations & 15& 15\\
    Mini-batch size & 100& 100\\
    Number of GRU layers, intra-session level & 1& 1\\
    Number of GRU layers, inter-session level & 1& 1\\
  \bottomrule
\end{tabular}
\caption{Best configurations for the RNN models.}
\label{tab:hyperparam}
\end{table}

\section{Results and Discussion}
\label{subsec:results}

\subsection{Effect of parameter \texorpdfstring{$\alpha$}{Lg}}

\begin{figure*}[!ht]
    \centering
    \begin{subfigure}[ht]{\textwidth}
        \centering
        \includegraphics[width=0.459\linewidth]{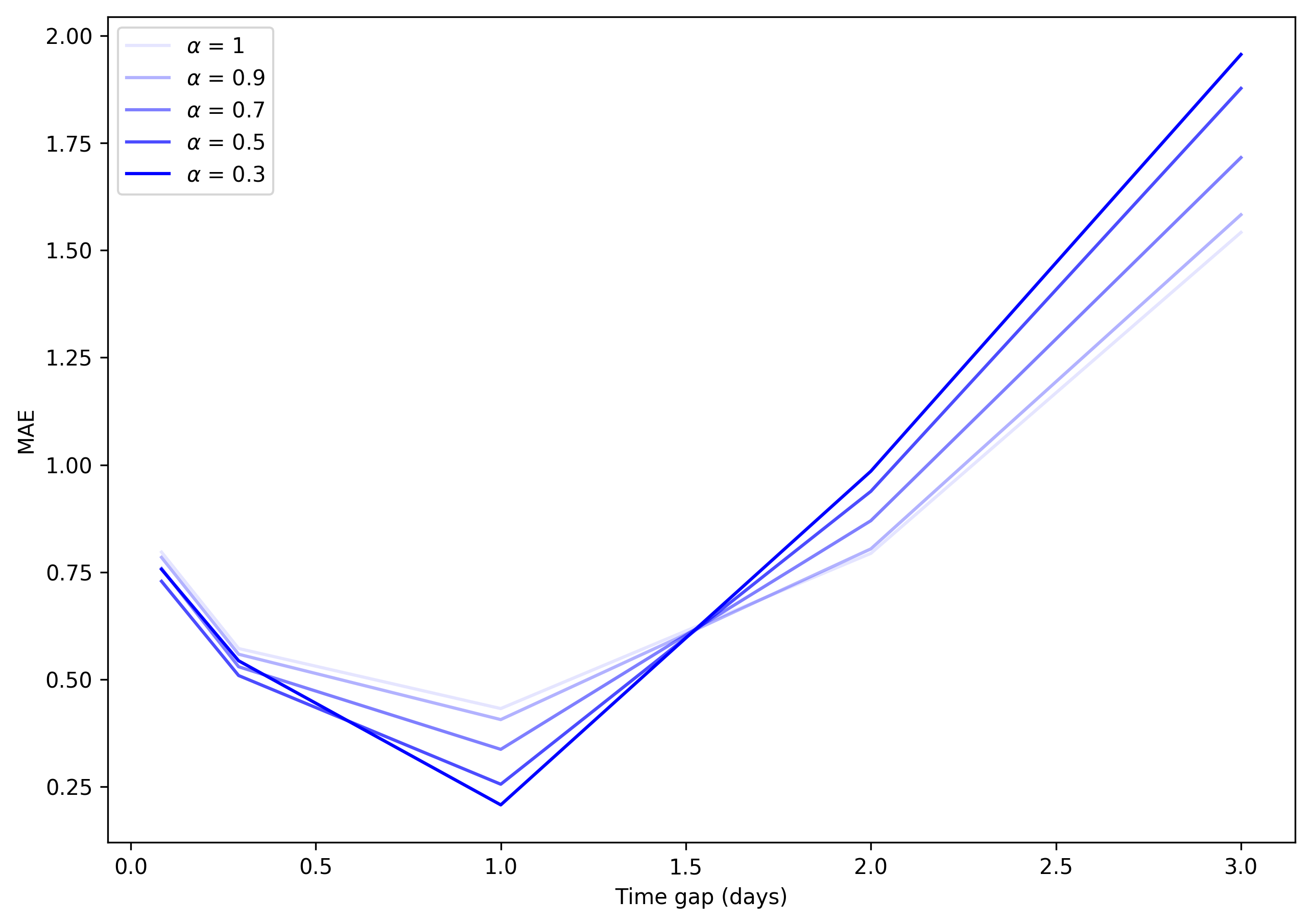}
        \hfill
        \includegraphics[width=0.450\linewidth]{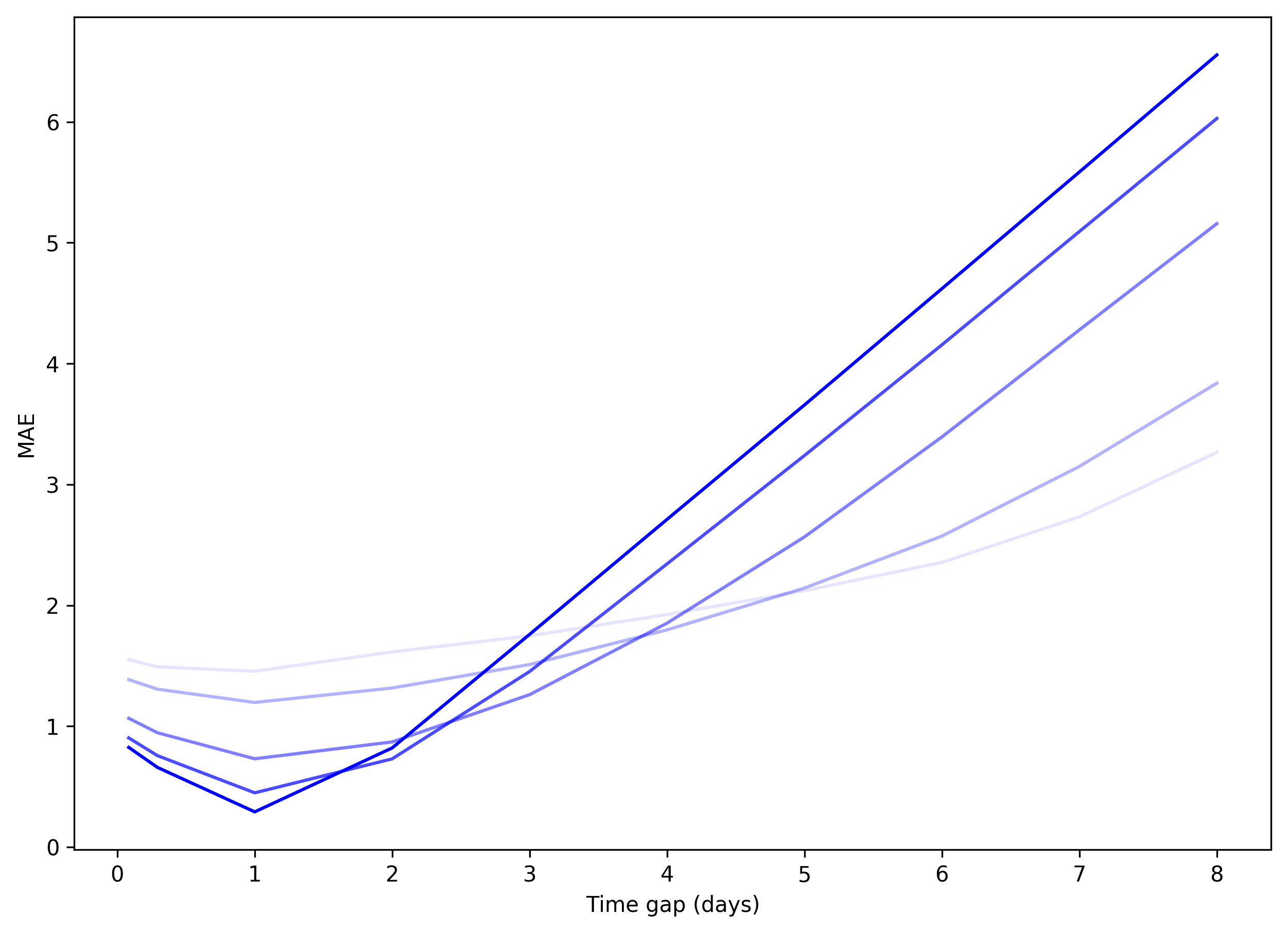}
    \end{subfigure}
    \caption{Time prediction MAE with different values of parameter$\alpha$ evaluated on LastFM dataset (left) and Reddit dataset (right). Runs with $\alpha = 0.1$ are not included since time-specific gradients diverged for this initialization}
    \label{fig:effect_alpha}
\end{figure*}

In order to show the effect of the parameter $\alpha$ in Equation \ref{eq:totloss}, we evaluate the model by varying $\alpha \in [0.3,0.5,0.7,0.9, 1.0]$ for the task of return time prediction. In Figure \ref{fig:effect_alpha} we see performance, in term of MAE, of the proposed architecture for both LastFM and Reddit dataset. In both cases we observe that with $\alpha$ close to $1.0$ the MAE decrease for longer time gap, showing an increase of accuracy in the time prediction for such gaps. 
In particular, for the LastFM dataset the performance of all the initializations seems to meet at $1.5$ days time-gaps which is the middle point between the $[0.5-1.5]$ days interval and $[1.5-2.5]$ days interval. We can also observe that the plot looks more linear the smaller $\alpha$ is set. For example, for $\alpha = 0.3$ the plot is similar to the one produced by predicting an averaged time-gap for every single prediction.
For the LastFM dataset, we see that by decreasing the value of $\alpha$, the focus of the model is to have better performance on smaller time-gaps but unlike the Reddit dataset, on the long-term gap, the performance of the model tends to deteriorate faster. This is probably due to differences in distribution of the time gaps for the two datasets.

For both datasets, Figure \ref{fig:effect_alpha} clearly illustrates the trade-off between error for the most frequent time-gaps and the effective prediction range of the model. Choosing $\alpha$ should be based on the service for which we are modeling return times. For some services, modeling shorter time gaps are more important that modeling longer ones.

\subsection{Return time prediction: Long Term vs Short Term}

\begin{figure*}[ht]
    \centering
    \begin{subfigure}[b]{\textwidth}
        \centering
        \includegraphics[width=0.33\linewidth]{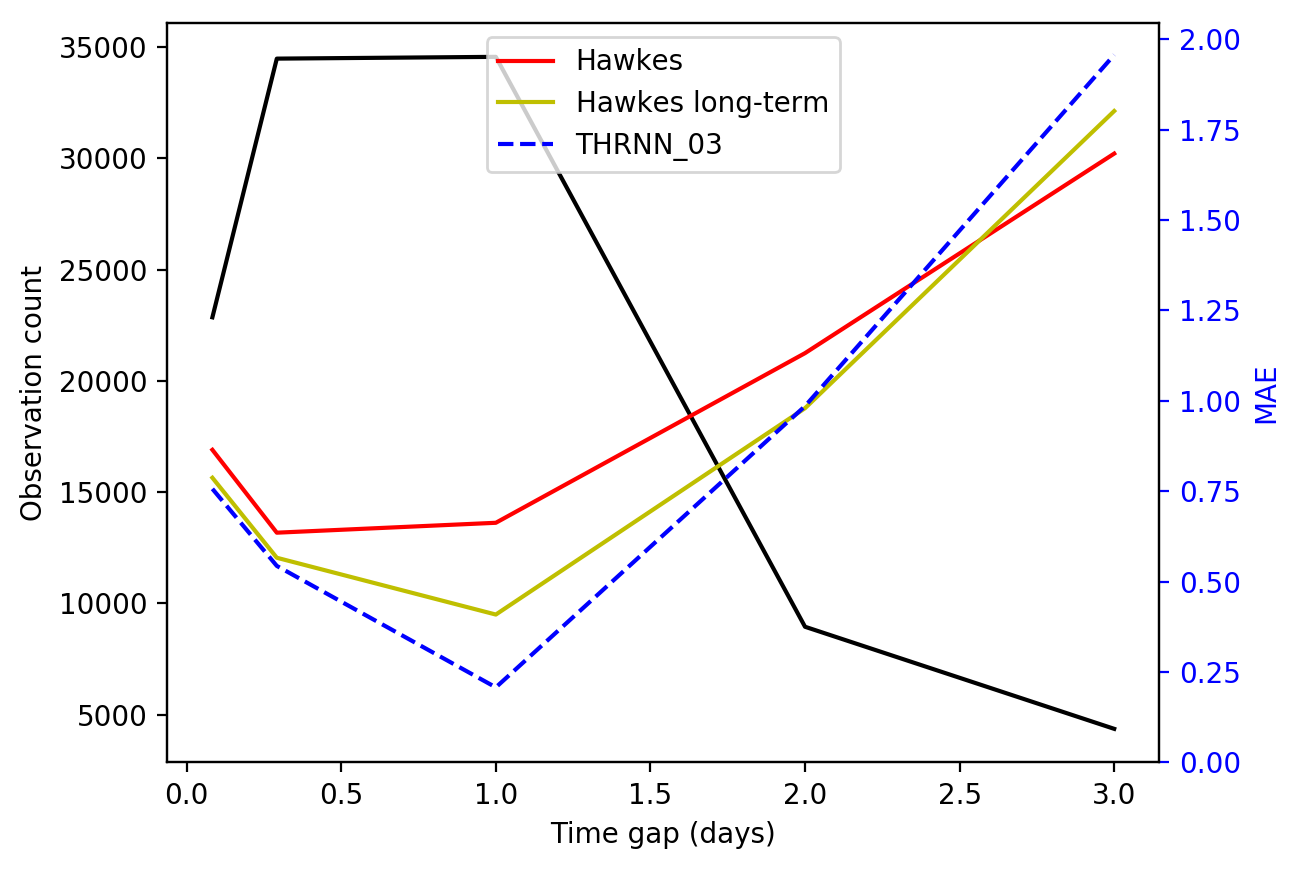}
        \hfill
        \includegraphics[width=0.33\linewidth]{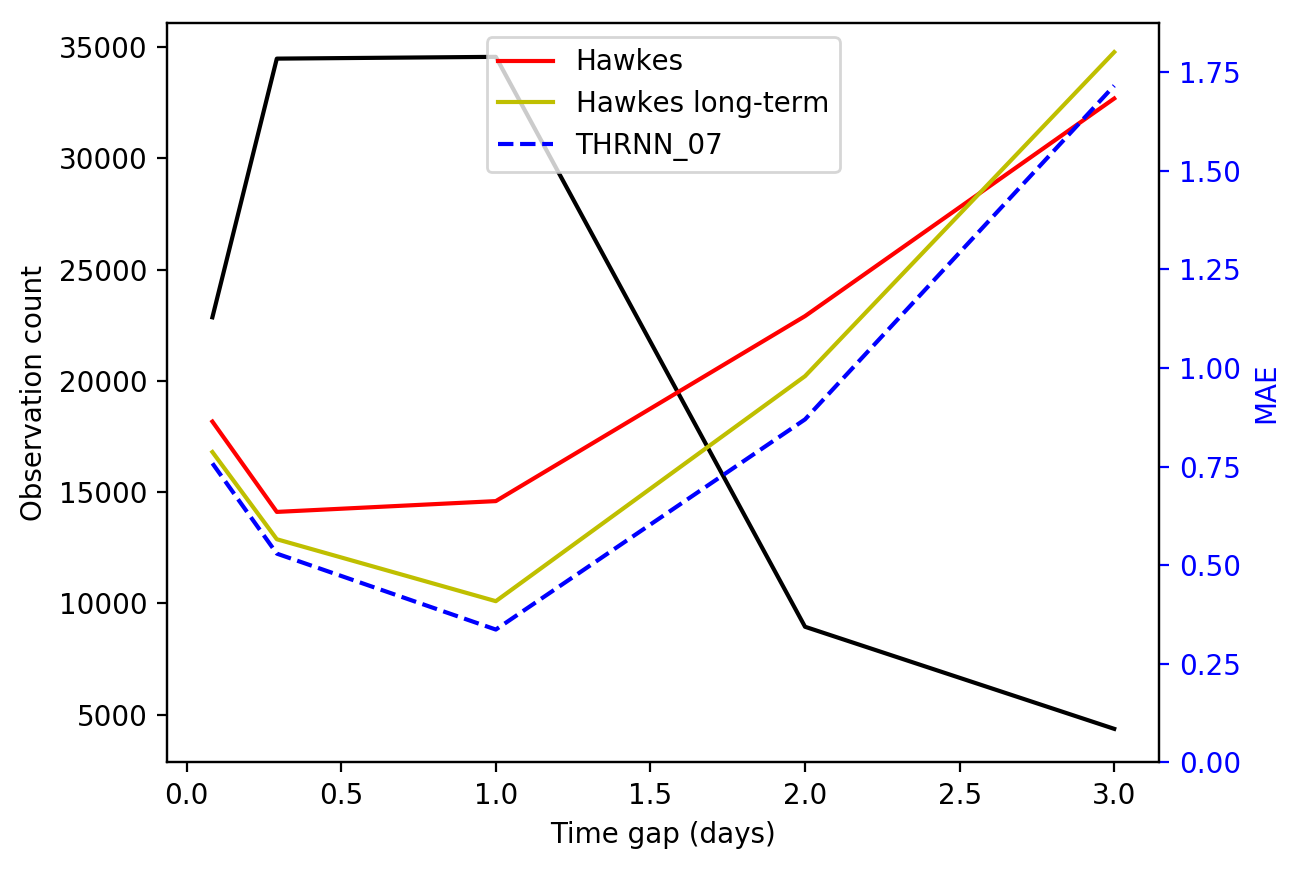}
        \hfill
        \includegraphics[width=0.33\linewidth]{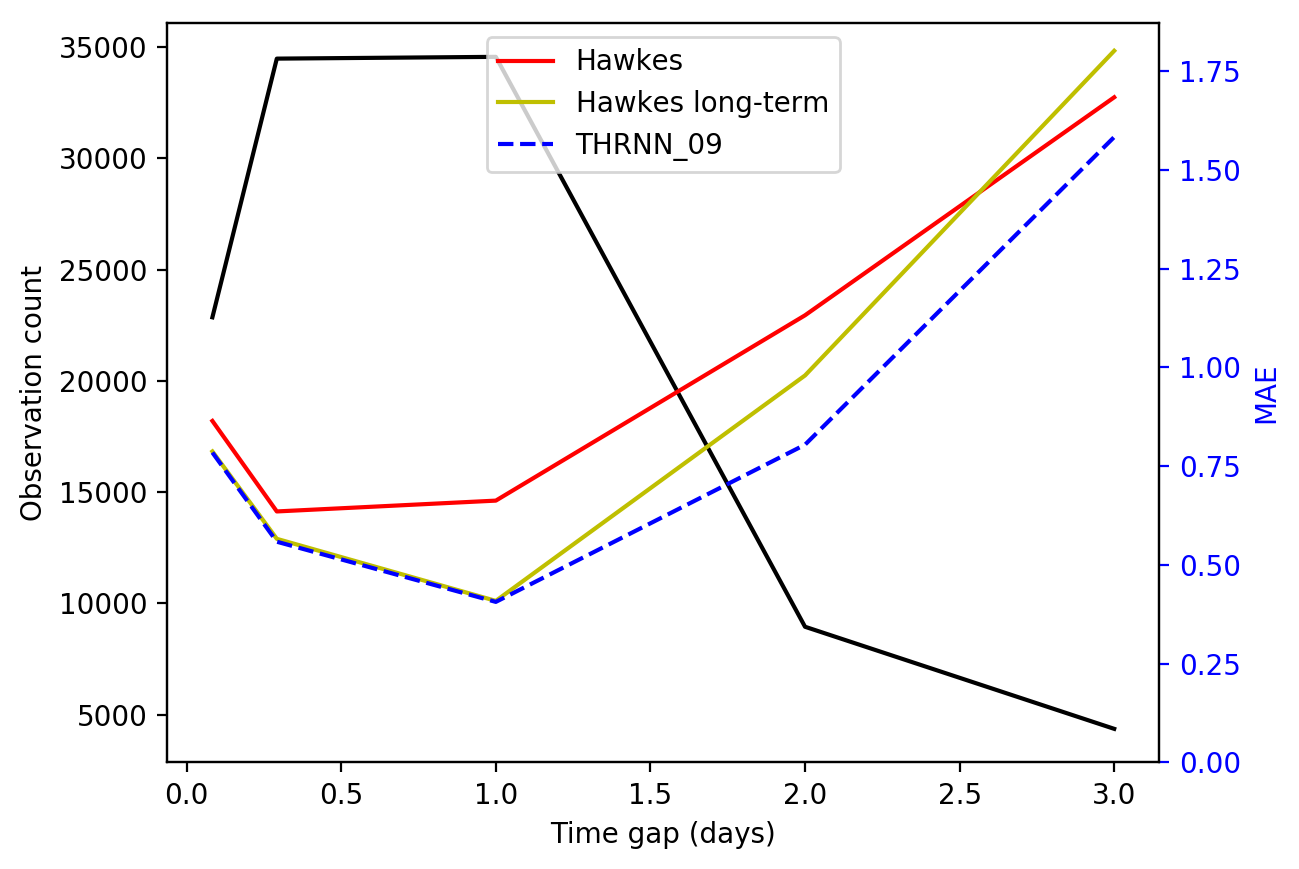}
        \caption{LastFM dataset: $\alpha$ values of $0.3$ (left), $0.5$ (center),  $0.9$ (right)}
    \end{subfigure}
    \vskip\baselineskip
    \begin{subfigure}[b]{\textwidth}
        \centering
        \includegraphics[width=0.33\linewidth]{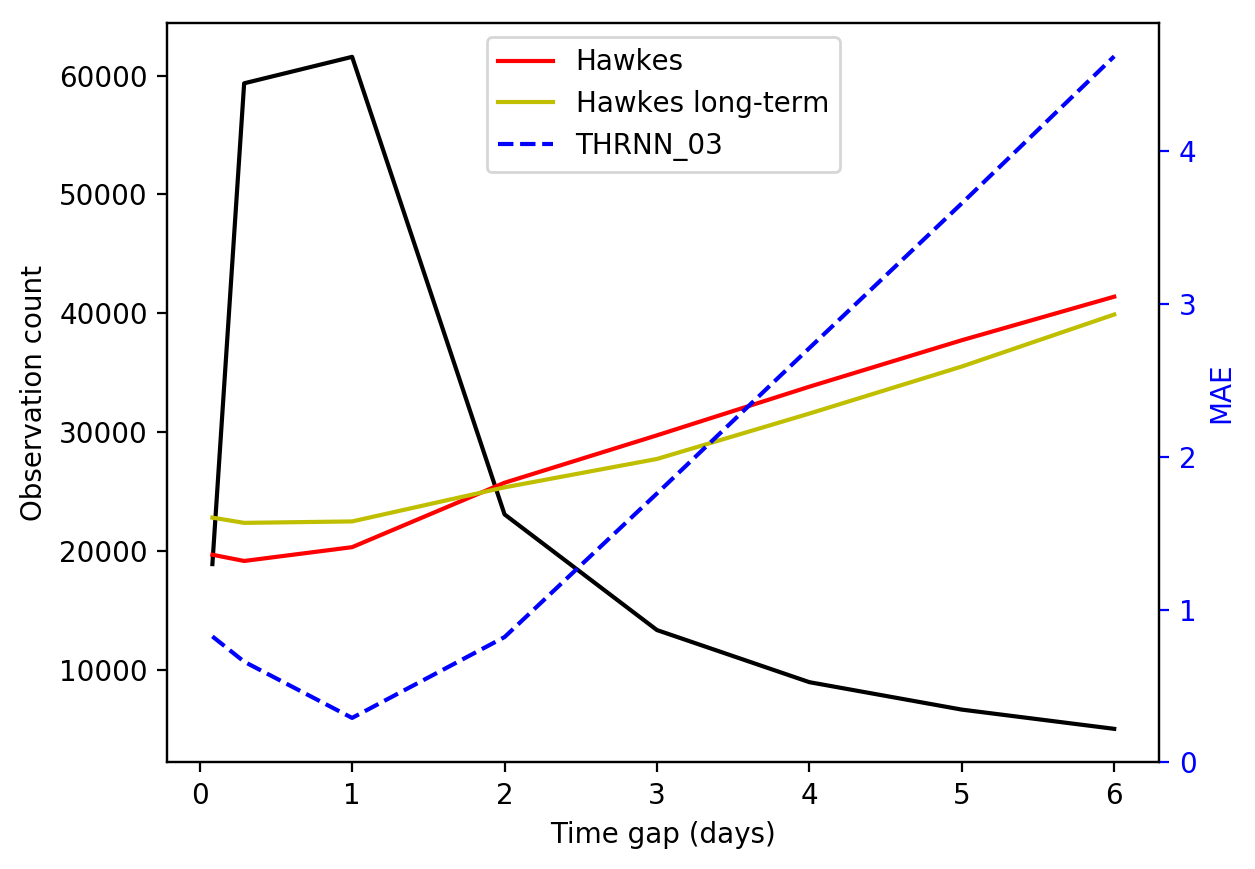}
        \hfill
        \includegraphics[width=0.33\linewidth]{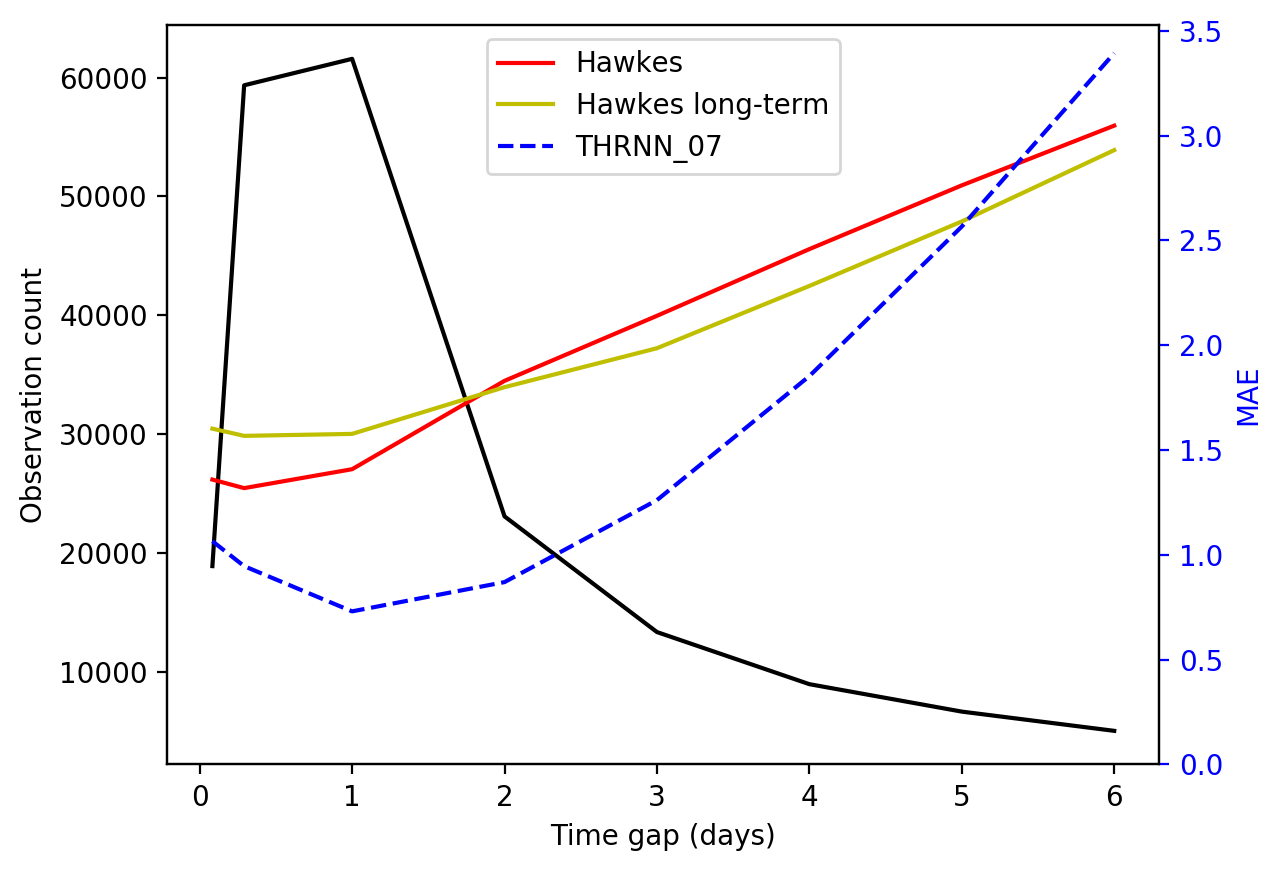}
        \hfill
        \includegraphics[width=0.33\linewidth]{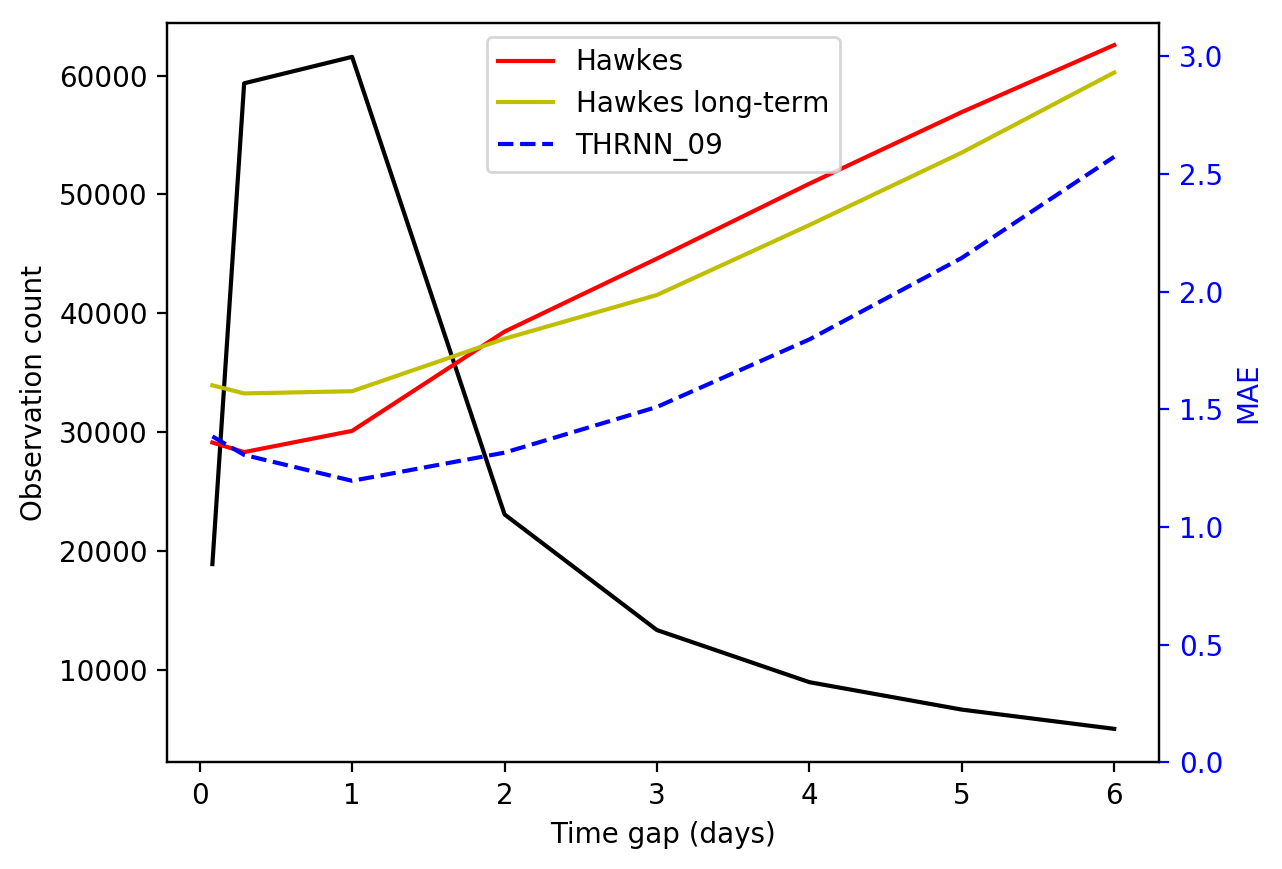}
        \caption{Reddit dataset: $\alpha$ values of $0.3$ (left), $0.5$ (center),  $0.9$ (right)}
    \end{subfigure}
    \caption{Time prediction MAE compared with Hawkes baselines with different values of parameter $\alpha$.}
    \label{fig:long_short}
\end{figure*}

In Figure \ref{fig:long_short}, we illustrate the performance of the proposed model for return time predictions versus two baseline Hawkes models, with tuning parameter $\alpha \in [0.3, 0.5, 0.9]$ in Equation \ref{eq:timeloss}. For short time-gaps, the joint model of session embeddings and point process intensity gives better predictions than the baselines. The results are plotted alongside the number of observations for each time-gap, which gives insightful information about how the user--item interaction within a session is relevant for the model. We observe that the joint model outperforms the baseline consistently at time-gaps where there are more user--item observations. This is intuitive because of the shared learned parameters between the time model and the intersession model, in the sense that those parameters are not only fitting the time-gaps, but also the user--item interactions dynamics. 

We see that tuning $\alpha$ allows us to control which time-gaps are most important to the model. Tuning for the most important time-gaps, however, comes at the cost of having worse return-time predictions for the time-gaps that are deemed less important.

\begin{table*}[ht]
\begin{tabular}{l l l l l l l}
 & \textbf{R@5} & \textbf{R@10} & \textbf{R@20} & \textbf{MRR@5} & \textbf{MRR@10} & \textbf{MRR@20} \\\hline
GRU4REC&$0.1349\pm0.0004$&$0.184\pm0.0002$&$0.2474\pm0.0002$&$0.086\pm0.0003$&$0.0925\pm0.0002$&$0.0969\pm0.0002$\\
\multirow{2}{*}{HRNN}&$0.1415\pm0.0005$&$0.1993\pm0.0007$&$0.2751\pm0.0006$&$0.0876\pm0.0004$&$0.0952\pm0.0004$&$0.1004\pm0.0004$\\
 &(+4.9\%)&(+8.3\%)&(+11.2\%)&(+1.8\%)&(+2.9\%)&(+3.7\%)\\
 \multirow{2}{*}{THRNN}& $\bm{0.1437\pm0.0004}$ &
 $\bm{0.2026\pm0.0006}$ &
 $\bm{0.2795\pm0.0006}$ &
 $\bm{0.0889\pm0.0002}$ &
 $\bm{0.0967\pm0.0003}$ &
 $\bm{0.102\pm0.0003}$\\
 &(+6.6\%)&(+10.1\%)&(+13.0\%)&(+3.4\%)&(+4.5\%)&(+5.3\%)\\
\end{tabular}
\caption[Intra-session recommendation results compared with RNN, LastFM]{Table with the recall and MRR on the LastFM dataset.}
\label{tab:results_lastfm}
\end{table*}

\begin{table*}[ht]
\begin{tabular}{l l l l l l l}
 & \textbf{R@5} & \textbf{R@10} & \textbf{R@20} & \textbf{MRR@5} & \textbf{MRR@10} & \textbf{MRR@20} \\\hline
GRU4REC&$0.3208\pm0.0004$&$0.3959\pm0.0005$&$0.475\pm0.0003$&$0.2346\pm0.0006$&$0.2445\pm0.0006$&$0.25\pm0.0006$\\
\multirow{2}{*}{HRNN}&$0.4432\pm0.0013$&$0.5316\pm0.0009$&$0.616\pm0.0012$&$0.317\pm0.0016$&$0.3288\pm0.0016$&$0.3347\pm0.0015$\\
 &(+38.1\%)&(+34.3\%)&(+29.7\%)&(+35.1\%)&(+34.5\%)&(+33.9\%)\\
 \multirow{2}{*}{THRNN}&
 $\bm{0.4468\pm0.0013}$ &
 $\bm{0.5366\pm0.001}$ &
 $\bm{0.6228\pm0.0009}$ &
 $\bm{0.3191\pm0.0015}$ &
 $\bm{0.3311\pm0.0014}$ &
 $\bm{0.3371\pm0.0014}$\\
 &(+39.3\%)&(+35.6\%)&(+31.1\%)&(+36.0\%)&(+35.4\%)&(+34.8\%)\\
\end{tabular}
\caption[Intra-session recommendation results compared with RNN, Reddit]{Table with the recall and MRR on the Reddit dataset.}
\label{tab:results_reddit}
\end{table*}

\subsection{Impact on actual recommendation}

The experimental results for the recommendation task are summarized on Table~\ref{tab:results_lastfm} (for the LastFm dataset) and Table~\ref{tab:results_reddit} (for the Reddit dataset). We observe an improvement in both metrics (Recall and MRR) at distinct levels (5, 10 and 20). This indicates that there is a win-win situation by incorporating intensity-based time modeling in the HRNN model, improving both recommendations and return time prediction when compared to respective baselines. However, it is worth noting that the difference between THRNN and HRNN is not just the time modeling, but also some added use of contexts in the THRNN model. It is also noticeable that the improvements of THRNN over HRNN, although significant, are less salient in absolute value than the improvements over GRU4REC. The simplest hypothesis is that the inter-session layer of HRNN is already capturing some of temporal dynamics between the sessions, for example it is possible that the latent representation are encoding time information correlated with the changes of user--item interactions from the end of one session to the beginning of the next session (for example if there are different profiles for distinct time-gaps of items that are typically accessed when finishing a session and items that typically are accessed in the beginning of the next session).

\section{Conclusion}
\label{sec:conclusion}
In this article we introduced a new joint model capable of intra-session and inter-session recommendations and inter-session return time prediction. Our results indicates that jointly modeling of intra-session and inter-session user-item interactions, both for recommendations and time-predictions is favorable for the session-based items recommendations and return time prediction task. The joint model is competitive with the state-of the art in both the recommendation and time-prediction task. Particularly in the latter we report significant improvement when compared to state-of-the-art models for time prediction in online social networks (Hawkes point process models). 
Also, we introduced in the loss function a mechanism for controlling and trade-off short, medium and long time prediction accuracy and show in the results that it is effective. Future studies within this framework could, e.g., entail how we can fuse multiple losses, each focusing on different time-gaps, for a more holistic loss.

\bibliographystyle{ACM-Reference-Format}
\bibliography{sigproc} 

\end{document}